\documentclass[11pt,draftcls,onecolumn]{IEEEtran}

\usepackage{amsmath}
\usepackage{graphicx}
\usepackage{epsfig}
\usepackage{array}
\usepackage{psfrag}
\usepackage{amssymb}
\usepackage{color}

\title{Filter-And-Forward Distributed Beamforming in Relay Networks with Frequency
Selective Fading}

\author{Haihua\ Chen, Alex\ B.\ Gershman, and Shahram Shahbazpanahi
\thanks{H.~Chen and A.~B.~Gershman
are with the Department of Communication Systems, Technische
Universit\"at Darmstadt, Merckstr.\ 25, 64283 Darmstadt, Germany;
emails: {\tt haihua.chen@nt.tu-darmstadt.de} and {\tt
gershman@nt.tu-darmstadt.de}, fax: +49 6151 162913, phone: +49
6151 162813. S.~Shahbazpanahi is with the Faculty of Engineering
and Applied Science, University of Ontario Institute of
Technology, 2000 Simcoe Street North, Oshawa, ON L1H7K4, Canada;
email: {\tt shahram.shahbazpanahi@uoit.ca},
fax: +1 905 721 3370, phone: +1 905 721 3111 ext: 2842. }
\thanks{A.~B.~Gershman is the corresponding
author.}
\thanks{This work was supported in parts by the European Research
Council (ERC) Advanced Investigator Grants program under Grant
227477-ROSE, German Research Foundation (DFG) under Grant GE
1881/1-1, and National Science and Engineering Research Council of
Canada (NSERC) under Discovery Grants program. The results of this
paper have been presented in part at IEEE/ITG Workshop on Smart
Antennas, Berlin, February 2009 and {\it ICASSP'09}, Taipei,
Taiwan, April 2009.}}

\begin{document}

\newlength{\figwidth}
\setlength{\figwidth}{3.3in}


\maketitle

\begin{abstract}
A new approach to distributed cooperative beamforming in relay
networks with frequency selective fading is proposed. It is
assumed that all the relay nodes are equipped with finite impulse
response (FIR) filters and use a filter-and-forward (FF) strategy
to compensate for the transmitter-to-relay and
relay-to-destination channels.

Three relevant half-duplex distributed beamforming problems are
considered. The first problem amounts to minimizing the total
relay transmitted power subject to the destination
quality-of-service (QoS) constraint. In the second and third
problems, the destination QoS is maximized subject to the total
and individual relay transmitted power constraints, respectively.
For the first and second problems, closed-form solutions are
obtained, whereas the third problem is solved using convex
optimization. The latter convex optimization technique can be also
directly extended to the case when the individual and total power
constraints should be jointly taken into account. Simulation
results demonstrate that in the frequency selective fading case,
the proposed FF approach provides substantial performance
improvements as compared to the commonly used amplify-and-forward
(AF) relay beamforming strategy.
\end{abstract}


\section{Introduction}
Recently, cooperative wireless communication techniques gained
much interest in the literature as they can exploit cooperative
diversity without any need of having multiple antennas at each
user \cite{laneman04}-\cite{nosratinia04a}. In such
user-cooperative schemes, different users share their
communication resources to assist each other in transmitting the
information throughout the network by means of relaying messages
from the source to destination through multiple independent paths.

Different relaying strategies have been proposed to achieve
cooperative diversity. Two most popular relaying strategies are
the amplify-and-forward (AF) and decode-and-forw\-a\-rd (DF)
approaches \cite{laneman04}, \cite{kramer05},
\cite{boelcskei04}-\cite{luoblum}. In the AF scheme, relays simply
retransmit properly scaled and phase-shifted versions of their
received signals, while in the DF scheme the relay nodes decode
and then re-encode their received messages prior to retransmitting
them. Due to its low complexity, the AF relaying strategy is of
especial interest \cite{jing_hass06}-\cite{fazeli08}.

When the channel state information (CSI) is not available at the
relay nodes, distributed space-time coding can be used to obtain
the cooperative diversity gain
\cite{laneman_wornell03}-\cite{jing_jaf08}. However, with
available CSI, distributed network beamforming can provide better
performance \cite{jing07}, \cite{jafar_netw_subm}.

Recently, several distributed AF beamforming techniques for relay
networks with flat fading channels have been developed
\cite{jing07}-\cite{fazeli08}. The approaches of
\cite{jing07}-\cite{zheng} optimize the receiver
quality-of-service (QoS) subject to the individual and/or total
power constraints under the assumption that the instantaneous CSI
is perfectly known at the destination or relay nodes.
In these approaches, the QoS is measured in terms
of the signal-to-noise ratio (SNR) at the receiver. In
\cite{havary08}, the problem of AF relay beamforming is considered
under the assumption that the second-order statistics of the
source-to-relay and relay-to-destination channels are available.
Based on the latter assumption, several distributed beamforming
algorithms are developed in \cite{havary08}. In the first
technique of \cite{havary08}, the total relay transmit power is
minimized subject to the receiver QoS constraint, whereas in the
second approach of \cite{havary08}, the receiver QoS is maximized
subject to the total or individual relay power constraints. In
\cite{fazeli08}, the approach of \cite{havary08} has been extended
to multiple source-destination pairs. Recently, the problem of
using an imperfect (e.g., quantized) CSI feedback in distributed
beamforming has been also considered \cite{koy08}.

Although some extensions of distributed space time-coding
techniques to the frequency selective fading case are known in the
literature \cite{mheidat}, the problem of distributed beamforming
in frequency selective environments has not been addressed so far.
In particular, all the techniques of \cite{jing07}-\cite{koy08}
assume the transmitter-to-relay and relay-to-destination channels
to be frequency flat. However, in practical scenarios these
channels are likely to be frequency selective. In the latter case,
there is a significant amount of inter-symbol interference (ISI)
which makes it difficult to directly extend the techniques of
\cite{jing07}-\cite{koy08} to frequency selective fading channel
scenarios.

In this paper, we consider a relay network of one transmitter, one
destination, and multiple relay nodes under the assumption of
frequency selective, finite impulse response (FIR)
transmitter-to-relay and relay-to-destination channels. To
compensate for the effect of such channels, a new
filter-and-forward (FF) relaying protocol is proposed. According
to the FF strategy, all the relay nodes are equipped with finite
impulse response (FIR) filters that are used to compensate for the
transmitter-to-relay and relay-to-destination channels.

Three relevant distributed beamforming problems are considered
under the assumption that the instantaneous CSI is available at
the receiver or at the relay nodes. Similar to the techniques of
\cite{jing07}-\cite{fazeli08}, the receiver is assumed to use a
perfect source-to-relay and relay-to-destination CSI to compute
the relay weight coefficients and feed them back to the relay
nodes using a low-rate receiver-to-relays feedback link.
Alternatively, the relays can directly compute their weight
coefficients, provided that the CSI is available at the relay
nodes.

Our first distributed beamforming problem amounts to minimizing
the total relay transmitted power subject to the destination QoS
constraint. As in the frequency selective case the
major performance limiting factor is ISI rather than noise, the
QoS is measured in terms of the receiver
signal-to-interference-plus-noise ratio (SINR), in contrast to the
techniques of \cite{jing07}-\cite{havary08} that use the SNR as a
measure of QoS in the flat fading case. In our
second and third problems, the destination QoS is maximized
subject to the total and individual relay transmitted power
constraints, respectively. For the first and second problems,
closed-form solutions are obtained, whereas the third problem is
solved using convex optimization. The latter convex optimization
technique can be also directly extended to the case when the
individual and total power constraints should be jointly taken
into account.

It is shown that in the flat fading case, the
proposed FF network beamforming techniques reduce to the AF
network beamformers of \cite{jing07}-\cite{havary08} which are
particular cases of our techniques.

In frequency selective channel scenarios, our simulation results
demonstrate that the proposed FF approach provides substantial
performance improvements as compared to the AF relay beamforming
strategy.

\section{Relay Network Model}

\subsection{Filter-And-Forward Relaying
Protocol} Let us consider a half-duplex relay network with one
single-antenna transmitting source, one single-antenna receiver
(destination) node and $R$ single-antenna relay nodes. Similar to
\cite{jing07} and \cite{zheng}-\cite{fazeli08}, it is assumed that
there is no direct link between the transmitter and destination
nodes and that the network is perfectly synchronized. Each
transmission is assumed to consist of two stages. In the first
stage, the transmitting source broadcasts its data to the relays.
The signals received at the relay nodes are then passed through
the relay FIR filters to compensate for the effects of the
transmitter-to-relay and relay-to-destination frequency selective
ch\-a\-nnels. This type of relay processing
corresponds to our proposed FF relaying protocol; see
Fig.~\ref{sys_model}. As FIR filters have been commonly used for
channel equalization in point-to-point communication systems, the
FF strategy appears to be a very natural extension of the AF
protocol to frequency selective relay channels. However, an
important difference between these two cases is that in relay
networks, it is meaningful to use a separate FIR filter at each
relay node, while in the traditional point-to-point case, such a
filter is commonly employed at the receiver.

In the second transmission stage, the outputs of each relay filter
are sent to the destination that is assumed to have the full
instantaneous CSI. Using this knowledge, the receiver determines
the filter weight coefficients of each relay according to a
certain beamforming criterion. It is also assumed that there is a
low-rate feedback link from the destination to each relay node
that is used to inform the relays about their optimal weight
coefficients. Alternatively, if the full instantaneous CSI is
available at the relays rather than the destination, each relay
node can determine its own weight coefficients
independently.{\footnote{Note that
network beamforming is commonly referred to as ``distributed''
because it is assumed that no relay can share its received signals
with any other relay.} In the latter case, no extra
receiver-to-relay feedback is needed. Note that quite similar
assumptions have been used in \cite{jing07}-\cite{fazeli08} in the
frequency flat fading case.

\subsection{Signal Model}

Let us model the transmitter-to-relay and relay-to-destination
channels as linear FIR filters
\begin{eqnarray}
{\mathbf f}(\omega) = \sum_{l=0}^{L_f-1} {\mathbf f}_l e^{-j\omega
l},\qquad {\mathbf g}(\omega) = \sum_{l=0}^{L_g-1} {\mathbf g}_l
e^{-j\omega l}
\end{eqnarray}
where
\begin{eqnarray}
{\mathbf f}_l&\!\!\!=\!\!\!& [f_{l,1},\ldots,
f_{l,R}]^T\label{cha_imp1}\\ {\mathbf
g}_l&\!\!\!=\!\!\!&[g_{l,1},\ldots, g_{l,R}]^T \label{cha_imp2}
\end{eqnarray}
are the $R\times 1$ channel impulse response vectors corresponding
to the $l$th effective tap of the transmitter-to-relay and
relay-to-destination channels, respectively. Here, ${\mathbf
f}(\omega)$ and ${\mathbf g}(\omega)$ are the $R\times 1$ vectors
of channel frequency responses, and $L_f$ and $L_g$ are the
corresponding channel lengths, respectively. The $R\times 1$
vector ${\mathbf r}(n)=[r_1(n),\cdots, r_R(n)]^T$ of the signals
received by the relay nodes in the $n$th channel use can be
modeled as
\begin{eqnarray}
{\mathbf r}(n) = \sum_{l=0}^{L_f-1} {\mathbf f}_l s(n-l) +
{\boldsymbol \eta} (n) \label{relay_sig}
\end{eqnarray}
where $s(n)$ is the signal transmitted by the source,
${\boldsymbol \eta} (n) = [\eta_1(n),\cdots,\eta_R(n)]^T$ is the
$R\times 1$ vector of relay noise, and $(\cdot)^T$ denotes the
transpose. Introducing the notations
\begin{eqnarray*}
{\mathbf F} &\!\!\!\triangleq\!\!\!& [{\mathbf
f}_0,\cdots,{\mathbf f}_{L_f-1}]\\
{\mathbf s} (n) &\!\!\!\triangleq\!\!\!&
[s(n),s(n-1),\cdots,s(n-L_f+1)]^T
\end{eqnarray*}
we can write (\ref{relay_sig}) as
\begin{eqnarray}
{\mathbf r}(n) = {\mathbf F} {\mathbf s}(n) + {\boldsymbol \eta}
(n) .\label{relay_sig_mtx}
\end{eqnarray}
The signal vector ${\mathbf t}(n)= [t_1(n),\cdots, t_R(n)]^T$ sent
from the relays to the destination can be expressed as
\begin{eqnarray}
{\mathbf t}(n) = \sum_{l=0}^{L_w-1} {\mathbf W}_l^H {\mathbf
r}(n-l) \label{relay_trans}
\end{eqnarray}
where $$ {\mathbf W}_l \triangleq {\rm diag}
\{w_{l,1},\cdots,w_{l,R}\}$$ is the diagonal matrix of the relay
filter impulse responses corresponding to the $l$th effective
filter tap of each relay, $L_w$ is the length of the relay FIR
filters, $(\cdot)^H$ denotes the Hermitian transpose, and for any
vector ${\boldsymbol x}$, the operator ${\rm diag}\{{\boldsymbol
x}\}$ forms the diagonal matrix containing the entries of
${\boldsymbol x}$ on its main diagonal. Correspondingly, for any
square matrix ${\boldsymbol X}$, the operator ${\rm
diag}\{{\boldsymbol X}\}$ forms a vector whose elements are the
diagonal entries of ${\boldsymbol X}$. Note that
if $L_w=1$, then the FF transmission in (\ref{relay_trans})
reduces to the AF one.

Inserting (\ref{relay_sig_mtx}) into (\ref{relay_trans}), we have
\begin{eqnarray}
{\mathbf t}(n) &\!\!\!=\!\!\!& \sum_{l=0}^{L_w-1} {\mathbf
W}_l^H({\mathbf F} {\mathbf s}(n-l) + {\boldsymbol \eta} (n-l)).
\label{relay_trans_2}
\end{eqnarray}
Let us define
\begin{eqnarray*}
 {\tilde {\mathbf s}}(n) &\!\!\!\!\triangleq\!\!\!\!& [s(n),s(n-1),\cdots,s(n-L_f-L_w+2)]^T.
\end{eqnarray*}
It can be seen that the vector ${\mathbf s}(n-l)$ is a subvector
of ${\tilde {\mathbf s}}(n)$. Using this observation,
\eqref{relay_trans_2} can be rewritten as
\begin{eqnarray}
{\mathbf t}(n)&\!\!\!=\!\!\!& \sum_{l=0}^{L_w-1} {\mathbf W}_l^H
({\mathbf F}_l {\tilde {\mathbf s}}(n)+{\boldsymbol \eta}
(n-l))\label{relay_trans1}
\end{eqnarray}
where
\begin{eqnarray*}
{\bf F}_l \triangleq [\,\overbrace{{\mathbf 0}_{R \times 1},
\cdots,{\mathbf 0}_{R \times 1}}^{l~{\rm columns}},\,{\bf F}, \,
\overbrace{{\mathbf 0}_{R \times 1},\cdots, {\mathbf 0}_{R \times
1}}^{(L_w-1-l)~{\rm columns} }\,], \ \  l = 0,\cdots,L_w-1.
\end{eqnarray*}
Let us also define
\begin{eqnarray*}
{\mathbf W} &\!\!\!\!\triangleq\!\!\!\!& [{\mathbf W}_0 ,\cdots,{\mathbf W}_{L_w-1}]^T\\
&&\hspace{18.5em}\overbrace{\hspace{6.7em}}^{L_w-1~{\rm columns}} \\
{\boldsymbol {\cal F}} &\!\!\!\!\triangleq\!\!\!\!& \left[
\begin{array}{c}
{\bf F}_0\\
{\bf F}_1\\
\vdots\\
{\bf F}_{L_w-1}
\end{array}
\right] = \left[\!
\begin{array}{ccccccc}
{\mathbf f}_0& {\mathbf f}_1 \!&\! \cdots \!&\! {\mathbf f}_{L_f-1} \!&\! {\mathbf 0}_{R\times 1}
\!&\! \cdots \!&\! {\mathbf 0}_{R\times 1} \\
{\mathbf 0}_{R\times 1}\! &\! {\mathbf f}_0 \!&\! {\mathbf f}_1
\!&\! \cdots \!&
\!{\mathbf f}_{L_f-1}\! &\!\cdots \!& \!{\mathbf 0}_{R\times 1} \\
\!&\!\! &\!\ddots\! &\! \ddots \!&\!\! &\!\ddots\! &\! \\
{\mathbf 0}_{R\times 1}\! &\! {\mathbf 0}_{R\times 1} \!&
\!\cdots\! &\! {\mathbf f}_0\!&\! {\mathbf f}_1\! &\! \cdots \!&
\!{\mathbf f}_{L_f-1}
\end{array} \!\!\right]\\
{\tilde {\boldsymbol \eta}}(n) &\!\!\!\!\triangleq\!\!\!\!&
[{\boldsymbol \eta}^T(n),{\boldsymbol \eta}^T(n-1), \cdots ,
{\boldsymbol \eta}^T (n-L_w+1)]^T
\end{eqnarray*}
where ${\mathbf 0}_{N\times M}$ is the $N \times M$ matrix of
zeros. Using these notations, we obtain that \begin{eqnarray*}
\sum_{l=0}^{L_w-1} {\mathbf W}_l^H {\mathbf F}_l = {\mathbf W}^H
{\boldsymbol {\cal F}}
\end{eqnarray*}
and, therefore, (\ref{relay_trans1}) can be expressed as
\begin{eqnarray}
{\mathbf t}(n) &\!\!\!=\!\!\!& {\mathbf W}^H {\boldsymbol {\cal
F}} {\tilde {\mathbf s}}(n) + {\mathbf W}^H {\tilde {\boldsymbol
\eta}}(n). \label{relay_trans_mtx}
\end{eqnarray}
The received signal at the destination can be
written as
\begin{eqnarray}
y(n) = \sum_{l=0}^{L_g-1} {\mathbf g}_l^T {\mathbf t}(n-l) +
\upsilon(n) \label{dest_sig}
\end{eqnarray}
where $\upsilon(n)$ is the receiver noise waveform
and ${\mathbf g}_l$ is the channel impulse
response vector defined in (\ref{cha_imp2}).

Using (\ref{relay_trans_mtx}), we can rewrite
(\ref{dest_sig}) as
\begin{eqnarray}
y(n) = \sum_{l=0}^{L_g-1} {\mathbf g}_l^T {\mathbf W}^H
{\boldsymbol {\cal F}} {\tilde {\mathbf s}}(n-l) +
\sum_{l=0}^{L_g-1} {\mathbf g}_l^T{\mathbf W}^H {\tilde
{\boldsymbol \eta}}(n-l) + \upsilon(n). \label{dest_sig2}
\end{eqnarray}
Taking into account that the matrices ${\mathbf
W}_l$ are all diagonal and using the properties of the Kronecker
matrix product, we obtain that
\begin{eqnarray}
{\mathbf g}_l^T {\mathbf W}^H &\!\!\!=\!\!\!& [{\mathbf
g}_l^T{\mathbf
W}_0^H,\!\cdots\!,{\mathbf g}_l^T{\mathbf W}_{L_w-1}^H] \nonumber \\
&\!\!\!=\!\!\!& [{\mathbf w}_0^H{\mathbf G}_l,\!\cdots\!,{\mathbf w}_{L_w-1}^H {\mathbf G}_l] \nonumber \\
&\!\!\!=\!\!\!&{\mathbf w}^H ( {\mathbf I}_{L_w} \otimes {\mathbf
G}_l ) \label{temp wg}
\end{eqnarray}
where
\begin{eqnarray*}
{\mathbf w} &\!\!\!\triangleq\!\!\!& [{\mathbf
w}_0^T,\!\cdots\!,{\mathbf w}_{L_w-1}^T]^T\\
{\mathbf w}_l &\!\!\!\triangleq\!\!\!& {\rm diag} \{{\mathbf
W}_l\}\\
{\mathbf G}_l &\!\!\!\triangleq\!\!\!& {\rm diag} \{{\mathbf
g}_l\}
\end{eqnarray*}
${\mathbf I}_{N}$ is the $N \times N$ identity matrix, and
$\otimes$ denotes the Kronecker product. Using
(\ref{temp wg}), we can further rewrite (\ref{dest_sig2}) as
\begin{eqnarray}
y(n) &\!\!\!=\!\!\!& \sum_{l=0}^{L_g-1} {\mathbf w}^H ( {\mathbf
I}_{L_w} \otimes {\mathbf G}_l ) {\boldsymbol {\cal F}} {\tilde
{\mathbf s}}(n-l) + \sum_{l=0}^{L_g-1} {\mathbf w}^H ( {\mathbf
I}_{L_w} \otimes {\mathbf G}_l ) {\tilde {\boldsymbol \eta}}(n-l)
+ \upsilon(n). \label{dest_sig_temp}
\end{eqnarray}
Defining
\begin{eqnarray*}
{\breve{\mathbf s}}(n) &\!\!\!\!\triangleq\!\!\!\!&
[s(n),s(n-1),\cdots,s(n-L_f-L_w-L_g+3)]^T\\
{\breve{\boldsymbol \eta}}(n)
&\!\!\!\!\triangleq\!\!\!\!&[{\boldsymbol \eta}^T(n),{\boldsymbol
\eta}^T(n-1), \cdots , {\boldsymbol \eta}^T(n-L_w-L_g+2)]^T
\end{eqnarray*}
we notice that ${\tilde{\mathbf s}}(n-l)$ and ${\tilde{\boldsymbol
\eta}}(n-l)$ are subvectors of ${\breve{\mathbf s}}(n)$ and
${\breve{\boldsymbol \eta}}(n)$, respectively. Therefore, we can
express (\ref{dest_sig_temp}) as
\begin{eqnarray}
y(n) &\!\!\!=\!\!\!& \sum_{l=0}^{L_g-1} {\mathbf w}^H ( {\mathbf
I}_{L_w} \otimes {\mathbf G}_l ) {\boldsymbol {\cal F}}_l {\breve
{\mathbf s}}(n) + \sum_{l=0}^{L_g-1} {\mathbf w}^H ( {\mathbf
I}_{L_w} \otimes {\mathbf G}_l ) {\breve {\mathbf I}}_l {\breve
{\boldsymbol \eta}}(n) + \upsilon(n) \label{dest_sig_temp1}
\end{eqnarray}
where
\begin{eqnarray*}
{\boldsymbol {\cal F}}_l &\!\!\!\!\triangleq\!\!\!\!&
[\,\overbrace{{\mathbf 0}_{RL_w \times 1}, \cdots,{\mathbf
0}_{RL_w \times 1}}^{l~{\rm columns}},\,{\boldsymbol {\cal F}}, \,
\overbrace{{\mathbf 0}_{RL_w \times 1},\cdots,
{\mathbf 0}_{RL_w \times 1}}^{(L_g-1-l)~{\rm columns} }\,],  \ \  l = 0,\cdots,L_g-1 \\
{\breve {\mathbf I}}_l &\!\!\!\!\triangleq\!\!\!\!&
[\,\overbrace{{\mathbf 0}_{RL_w \times R},\!\cdots\!, {\mathbf
0}_{RL_w \times R}}^{l~{\rm blocks}},{\mathbf I}_{RL_w},
\overbrace{{\mathbf 0}_{RL_w \times R},\!\cdots\!,{\mathbf
0}_{RL_w \times R}}^{(L_g-1-l)~{\rm blocks}}\,],\ \  l =
0,\cdots,L_g-1.
\end{eqnarray*}

To express (\ref{dest_sig_temp1}) in a more compact form, we
further define
\begin{eqnarray}
{\boldsymbol {\cal G}} &\!\!\!\!\triangleq\!\!\!\!& \left[{\mathbf
I}_{L_w} \otimes {\mathbf G}_0,\cdots, {\mathbf I}_{L_w} \otimes
{\mathbf G}_{L_g-1}  \right] \nonumber\\
{\breve{\mathbf F}} &\!\!\!\!\triangleq\!\!\!\!& [ {\boldsymbol
{\cal F}}_0^T,
\cdots, {\boldsymbol {\cal F}}_{L_g-1}^T ]^T  \nonumber \\
{\tilde{\mathbf I}} &\!\!\!\!\triangleq\!\!\!\!& [{\breve {\mathbf
I}}_0^T,\cdots, {\breve {\mathbf I}}_{L_g-1}^T]^T \nonumber
\end{eqnarray}
and note that
\begin{eqnarray*}
\sum_{l=0}^{L_g-1} ({\mathbf I}_{L_w} \otimes {\mathbf G}_l )
{\boldsymbol {\cal F}}_l &\!\!\!=\!\!\!& {\boldsymbol{\cal G}}
{\breve{\mathbf F}}\\
\sum_{l=0}^{L_g-1} ( {\mathbf I}_{L_w} \otimes {\mathbf G}_l )
{\breve {\mathbf I}}_l &\!\!\!=\!\!\!& {\boldsymbol {\cal G}}
{\tilde{\mathbf I}}.
\end{eqnarray*}
Using the latter two equations, (\ref{dest_sig_temp1}) can be
expressed as
\begin{eqnarray}
y(n) = {\mathbf w}^H {\boldsymbol{\cal G}} {\breve{\mathbf F}}
{\breve{\mathbf s}}(n) + {\mathbf w}^H {\boldsymbol {\cal G}}
{\tilde{\mathbf I}} {\breve{\boldsymbol \eta}}(n) + \upsilon (n).
\label{dest_sig_temp2}
\end{eqnarray}
Let ${\bar{\mathbf f}}$ and ${\bar{\mathbf F}}$ denote the first
column and the residue of ${\breve{\mathbf F}}$, respectively, so
that ${\breve{\mathbf F}}= [{\bar{\mathbf f}}, {\bar{{\mathbf
F}}}]$. Then, (\ref{dest_sig_temp2}) yields
\begin{eqnarray}
y(n) &\!\!\!\!=\!\!\!\!& {\mathbf w}^H {\boldsymbol {\cal G}}
[{\bar{\mathbf f}}, {\bar{{\mathbf F}}}] \left[
\begin{array}{l}s(n)\\{\bar {\mathbf s}}(n) \end{array}\right] +
{\mathbf w}^H {\boldsymbol {\cal G}} {\tilde{\mathbf I}}
{\breve{\boldsymbol \eta}}(n)
+ \upsilon (n) \nonumber \\
&\!\!\!\!=\!\!\!\!& \underbrace{ {\mathbf w}^H {\boldsymbol {\cal
G}} {\bar{\mathbf f}} s(n) }_{ \rm signal} + \underbrace{ {\mathbf
w}^H {\boldsymbol {\cal G}} {\bar{{\mathbf F}}} {\bar{\mathbf
s}}(n)}_{\rm ISI} + \underbrace{ {\mathbf w}^H {\boldsymbol {\cal
G}} {\tilde{\mathbf I}} {\breve{\boldsymbol \eta}}(n) + \upsilon
(n) }_{\rm noise}\ \ \ \quad \label{dest_sig_mtx}
\end{eqnarray}
where $${\bar {\mathbf s}}(n) \triangleq
[s(n-1),\cdots,s(n-L_f-L_w-L_g+3)]^T.$$ In (\ref{dest_sig_mtx}),
we can identify the three components
\begin{eqnarray}
y_s (n) &\!\!\!\triangleq\!\!\!& {\mathbf w}^H {\boldsymbol {\cal G}} {\bar{\mathbf f}} s(n) \label{ys_t}\\
y_i (n) &\!\!\!\triangleq\!\!\!& {\mathbf w}^H {\boldsymbol {\cal G}} {{\bar{\mathbf F}}} {\bar{\mathbf s}}(n) \label{y_i}\\
y_n (n) &\!\!\!\triangleq\!\!\!& {\mathbf w}^H {\boldsymbol {\cal
G}} {\tilde{\mathbf I}} {\breve{\boldsymbol \eta}}(n) + \upsilon
(n) \label{y_n}
\end{eqnarray}
as the destination signal, ISI, and noise components,
respectively. Note that for the sake of computational simplicity
of our techniques developed in the next section, {\it block
processing} is not considered here, that is, the signal copies
delayed by multipath are not coherently combined.

The signal component in (\ref{ys_t}) can be expressed as
\begin{eqnarray}
y_s (n) &=& {\mathbf w}_0^H {\mathbf G}_0 {\mathbf f}_0 s(n) \nonumber\\
&=& {\mathbf w}_0^H ({\mathbf g}_0 \odot {\mathbf f}_0) s(n) \nonumber \\
&=& {\mathbf w}_0^H {\mathbf h}_0 s(n) \label{ys}
\end{eqnarray}
where
\begin{equation}
{\mathbf h}_0 \triangleq {\mathbf g}_0 \odot {\mathbf f}_0
\end{equation}
and $\odot$ denotes the Schur-Hadamard (elementwise) matrix
product.

\section{Filter-and-Forward Relay Beamforming} \label{formulation}
In this section, we develop three distributed FF beamforming
approaches that utilize several alternative
criteria. Our first FF beamforming technique is based on
minimizing the total relay transmitted power subject to the
destination QoS constraint, while our second and third approaches
are based on maximizing the destination QoS subject to the total
and individual relay transmitted power constraints, respectively.
A useful modification of our third approach is also discussed,
that enables to combine the later two types of constraints.

\subsection{Minimization of the Total Relay Power Under the QoS
Constraint}\label{subsec1} We first consider the distributed FF
beamforming problem that obtains the relay filter weights by
minimizing the total relay transmitted power $P$ subject to the
destination QoS constraint. As mentioned above, the destination
QoS is given by the receiver SINR
value\footnote{This is true
because the processing at the destination is rather simple; in
particular, no block processing is used.} and, therefore, the
latter problem can be written as
\begin{eqnarray}
\min_{{\mathbf w}} P \qquad {\rm s.t.} \quad {\rm SINR} \geq
\gamma \label{prob_orig}
\end{eqnarray}
where $\gamma$ is the minimal required SINR at the destination.

Let us use the following two common assumptions
\begin{eqnarray}
{\rm E}\{ {\tilde {\mathbf s}}(n) {\tilde {\mathbf s}}^H(n)\}=P_s
{\bf I}_{L_f+L_w-1},\quad {\rm E}\{{\tilde {\boldsymbol \eta}}(n)
{\tilde {\boldsymbol \eta}}^H(n)\}&\!\!\!=\!\!\!& \sigma_{\eta}^2
{\bf I}_{RL_w}\label{assum_white}
\end{eqnarray}
on statistical independence of the signal and
noise waveforms, respectively. Here, $P_s$ is the source
transmitted power and $\sigma_{\eta}^2$ is the relay noise
variance. Using (\ref{relay_trans_mtx}) and (\ref{assum_white}),
the transmitted power of the $m$th relay can be written as
\begin{eqnarray}
p_m &\!\!\!=\!\!\!& {\rm E} \{ \vert t_m(n)\vert^2\} \nonumber \\
&\!\!\!=\!\!\!& {\rm E} \{{\mathbf e}_m^T {\mathbf W}^H
{\boldsymbol {\cal F}} {\tilde {\mathbf s}}(n) {\tilde {\mathbf
s}}^H(n) {\boldsymbol {\cal F}}^H {\mathbf W} {\mathbf e}_m \} +
{\rm E} \{ {\mathbf e}_m^T {\mathbf W}^H {\tilde {\boldsymbol
\eta}}(n) {\tilde {\boldsymbol \eta}}^H(n)
{\mathbf W} {\mathbf e}_m \} \nonumber \\
&\!\!\!=\!\!\!& P_s {\mathbf e}_m^T {\mathbf W}^H {\boldsymbol
{\cal F}} {\boldsymbol {\cal F}}^H {\mathbf W} {\mathbf e}_m +
\sigma_{\eta}^2 {\mathbf e}_m^T {\mathbf W}^H {\mathbf W} {\mathbf
e}_m \label{p_i_W}
\end{eqnarray}
where ${\mathbf e}_m$ is the $m$th column of the identity matrix.

Using ${\mathbf E}_m\triangleq {\rm diag}\{{\mathbf e}_m\}$ and
the properties of the Kronecker product, (\ref{p_i_W}) can be
rewritten as
\begin{eqnarray}
p_m &\!\!\!=\!\!\!& P_s {\mathbf w}^H \left( {\mathbf I}_{L_w}
\otimes {\mathbf E}_m \right) {\boldsymbol {\cal F}} {\boldsymbol
{\cal F}}^H \left( {\mathbf I}_{L_w}
\otimes {\mathbf E}_m \right)^H {\mathbf w} \nonumber \\
&\!\!\!\!\!\!& + \sigma_{\eta}^2 {\mathbf w}^H \left( {\mathbf
I}_{L_w} \otimes {\mathbf E}_m \right) \left( {\mathbf I}_{L_w}
\otimes {\mathbf E}_m \right)^H {\mathbf w}. \label{p_i_w}
\end{eqnarray}
The total relay transmitted power can be then expressed as
\begin{eqnarray}
P = \sum_{m=1}^R p_m = {\mathbf w}^H \left(\sum_{m=1}^R {\mathbf
D}_m\right) {\mathbf w} = {\mathbf w}^H {\mathbf D} {\mathbf w}
\label{P_t}
\end{eqnarray}
where
\begin{eqnarray*}
{\mathbf D}_m &\!\!\!\triangleq \!\!\!& P_s  \left( {\mathbf
I}_{L_w} \otimes {\mathbf E}_m \right) {\boldsymbol {\cal F}}
{\boldsymbol {\cal F}}^H \left( {\mathbf I}_{L_w} \otimes {\mathbf
E}_m \right)^H + \sigma_{\eta}^2  \left( {\mathbf I}_{L_w} \otimes
{\mathbf E}_m \right) \left( {\mathbf I}_{L_w}
\otimes {\mathbf E}_m \right)^H \\
{\mathbf D} &\!\!\!\triangleq \!\!\!& \sum_{m=1}^R {\mathbf D}_m =
P_s \sum_{m=1}^R \left( {\mathbf I}_{L_w} \otimes {\mathbf E}_m
\right) {\boldsymbol {\cal F}} {\boldsymbol {\cal F}}^H \left(
{\mathbf I}_{L_w} \otimes {\mathbf E}_m \right)^H +
\sigma_{\eta}^2 {\mathbf I}_{RL_w}.
\end{eqnarray*}

The SINR at the
destination can be written as
\begin{equation}
{\rm SINR} = \frac{{\rm E}\{\vert y_s(n)\vert^2 \}}{{\rm E}\{\vert
y_i(n) \vert^2\} + {\rm E}\{\vert y_n(n) \vert^2 \}} .\label{SINR}
\end{equation}
Using (\ref{ys}), we obtain that
\begin{eqnarray}
{\rm E}\{\vert y_s(n)\vert^2 \} &\!\!\!=\!\!\!& {\rm E} \{
\vert {\mathbf w}_0^H {\mathbf h}_0 s(n)\vert^2\} \nonumber \\
&\!\!\!=\!\!\!& P_s {\mathbf w}_0^H {\mathbf h}_0 {\mathbf h}_0^H
{\mathbf w}_0\nonumber\\
&\!\!\!=\!\!\!& P_s {\mathbf w}^H {\mathbf A}^H {\mathbf h}_0 {\mathbf h}_0^H
{\mathbf A} {\mathbf w} \nonumber \\
&\!\!\!=\!\!\!& {\mathbf w}^H {\mathbf Q}_s {\mathbf w}
\label{P_s}
\end{eqnarray}
where
\begin{eqnarray*}
{\mathbf A} &\!\!\!\triangleq\!\!\!& [{\mathbf I}_R, {\mathbf
0}_{R\times (L_w-1)R}]\\
{\mathbf Q}_s &\!\!\!\triangleq\!\!\!& P_s {\mathbf A}^H {\mathbf
h}_0 {\mathbf h}_0^H {\mathbf A}.
\end{eqnarray*}

Using (\ref{y_i}), we have
\begin{eqnarray}
{\rm E}\{\vert y_i(n)\vert^2 \} &\!\!\!=\!\!\!& {\rm E}\{ {\mathbf
w}^H {\boldsymbol {\cal G}} {\bar{\mathbf F}} {\bar{\mathbf s}}(n)
{\bar{\mathbf s}}^H(n) {\bar{\mathbf F}}^H
{\boldsymbol{\cal G}}^H {\mathbf w}\} \nonumber \\
&\!\!\!=\!\!\!& P_s {\mathbf w}^H {\boldsymbol{\cal G}} {\bar{\mathbf F}}
{\bar{\mathbf F}}^H  {\boldsymbol{\cal G}}^H {\mathbf w} \nonumber \\
&\!\!\!=\!\!\!& {\mathbf w}^H {\mathbf Q}_i {\mathbf w}
\label{P_i}
\end{eqnarray}
where
\begin{eqnarray*}{\mathbf Q}_i \triangleq P_s {\boldsymbol {\cal G}}
{\bar{\mathbf F}} {\bar{\mathbf F}}^H  {\boldsymbol{\cal G}}^H.
\end{eqnarray*}

Making use of (\ref{y_n}), we also obtain that
\begin{eqnarray}
{\rm E}\{\vert y_n(n)\vert^2 \} &\!\!\!=\!\!\!& {\rm E}\{ {\mathbf
w}^H {\boldsymbol {\cal G}} {\tilde{\mathbf I}}
{\tilde{\boldsymbol \eta}}(n) {\tilde{\boldsymbol \eta}}^H(n)
{\tilde{\mathbf I}}^H
{\boldsymbol {\cal G}}^H {\mathbf w}\} + \sigma_{\upsilon}^2 \nonumber \\
&\!\!\!=\!\!\!& \sigma_{\eta}^2 {\mathbf w}^H {\boldsymbol {\cal
G}} {\tilde{\mathbf I}}\, {\tilde {\mathbf I}}^H {\boldsymbol
{\cal G}}^H {\mathbf w} +
\sigma_{\upsilon}^2 \nonumber \\
&\!\!\!=\!\!\!& {\mathbf w}^H {\mathbf Q}_n {\mathbf w} +
\sigma_{\upsilon}^2 \label{P_n}
\end{eqnarray}
where
\begin{eqnarray*}
{\mathbf Q}_n \triangleq \sigma_{\eta}^2 {\boldsymbol{\cal G}}
{\tilde{\mathbf I}}\, {\tilde{\mathbf I}}^H {\boldsymbol {\cal
G}}^H.
\end{eqnarray*}

Using (\ref{P_t}) and (\ref{P_s})-(\ref{P_n}), the problem in
(\ref{prob_orig}) can be rewritten in the following form:
\begin{eqnarray}
\min_{{\mathbf w}} \; {\mathbf w}^H {\mathbf D} {\mathbf w} \quad
{\rm s.t.} \quad \frac{{\mathbf w}^H {\mathbf Q}_s {\mathbf
w}}{{\mathbf w}^H {\mathbf Q}_i {\mathbf w} + {\mathbf w}^H
{\mathbf Q}_n {\mathbf w} + \sigma_{\upsilon}^2} \geq \gamma.
\label{prob_orig_sinr}
\end{eqnarray}
Introducing
\begin{equation}
\tilde {\mathbf w} \triangleq {\mathbf D}^{1/2} {\mathbf w}, \quad
{\mathbf Q} \triangleq {\mathbf D}^{-1/2} ({\mathbf Q}_s -
\gamma{\mathbf Q}_i - \gamma{\mathbf Q}_n) {\mathbf D}^{-1/2}
\end{equation}
we can reformulate the problem in
(\ref{prob_orig_sinr}) as
\begin{eqnarray}
\min_{\tilde{\mathbf w}} \; \|\tilde{\mathbf w}\|^2 \qquad {\rm s.t.} \quad
\tilde{\mathbf w}^H  {\mathbf Q} \tilde{\mathbf w} \geq  \gamma \sigma_{\upsilon}^2.
\label{prob_ineq}
\end{eqnarray}
The constraint function in (\ref{prob_ineq}) can be used for
checking the feasibility of the problem for any given value of
$\gamma$. In particular, for all the values of $\gamma$ that lead
to {\it negative semidefinite} ${\mathbf Q}$, the problem in
(\ref{prob_ineq}) is infeasible. It can be also easily proved that
the constraint in (\ref{prob_ineq}) can be replaced by the
equality constraint $\tilde{\mathbf w}^H {\mathbf Q}
\tilde{\mathbf w} = \gamma \sigma_{\upsilon}^2$. Hence, the
problem (\ref{prob_ineq}) is equivalent to
\begin{eqnarray}
\min_{\tilde{\mathbf w}} \; \|\tilde{\mathbf w}\|^2 \qquad {\rm
s.t.} \quad \tilde{\mathbf w}^H  {\mathbf Q} \tilde{\mathbf w} =
\gamma \sigma_{\upsilon}^2. \label{prob_eq}
\end{eqnarray}
The solution of (\ref{prob_eq}) can be found by means of
the Lagrange multiplier method. Let us minimize
the Lagrangian
\begin{equation}
H(\tilde{\mathbf w},\lambda) = \tilde{\mathbf w}^H \tilde{\mathbf w} +
\lambda (\gamma \sigma_{\upsilon}^2-\tilde{\mathbf w}^H  {\mathbf Q} \tilde{\mathbf w})
\label{Lagrange_fun}
\end{equation}
where $\lambda$ is a Lagrange multiplier. Taking gradient of
(\ref{Lagrange_fun}) and equating it to zero, we obtain that the
solution is equal to that of the following eigenvalue problem:
\begin{equation}
 {\mathbf Q} \tilde{\mathbf w}= \frac{1}{\lambda}\tilde{\mathbf
 w}.
\label{eigenvalue}
\end{equation}
Multiplying both sides of (\ref{eigenvalue}) with $\lambda
\tilde{\mathbf w}^H$ yields
\begin{equation}
\|\tilde{\mathbf w}\|^2=\tilde{\mathbf w}^H \tilde{\mathbf w}=\lambda
\tilde{\mathbf w}^H  {\mathbf Q} \tilde{\mathbf w} = \lambda \gamma \sigma_{\upsilon}^2.
\label{obj_eigen}
\end{equation}
It can be seen from (\ref{obj_eigen}) that minimizing
$\|\tilde{\mathbf w}\|^2$ leads to the smallest positive
$\lambda$, which is equivalent to the largest $1/\lambda$ in
(\ref{eigenvalue}). Using the latter fact, we conclude that the
optimal solution to (\ref{prob_ineq}) can be written as
\begin{eqnarray}
\tilde{\mathbf w}_{\rm opt} = \beta\, {\cal P} \{ {\mathbf Q} \}
\label{sol_w_tilde}
\end{eqnarray}
where ${\cal P} \{ {\cdot} \}$ denotes the normalized principal
eigenvector of a matrix and
\begin{eqnarray}
\beta=\left(\frac{\gamma \sigma_{\upsilon}^2}{ {\cal P} \{
{\mathbf Q} \}^H {\mathbf Q}\, {\cal P} \{ {\mathbf Q}
\}}\right)^{1/2}.
\end{eqnarray}

Therefore, the optimal beamformer weight vector and the minimum
total relay transmitted power can be expressed as
\begin{eqnarray}
{\mathbf w}_{\rm opt} &\!\!\!=\!\!\!& \beta\, {\mathbf D}^{-1/2}
{\cal P}
\{ {\mathbf Q} \} \label{sol_w} \\
P_{\rm min} &\!\!\!=\!\!\!& \gamma \sigma_{\upsilon}^2/{\cal
L}_{\rm max} \{{\mathbf Q}\}
\end{eqnarray}
respectively, where ${\cal L}_{\rm max} \{\cdot\}$ denotes the
largest (principal) eigenvalue of a matrix.

Hence, the FF distributed beamforming problem (\ref{prob_orig})
enjoys a simple closed-form solution based on the principal
eigenvector of the matrix ${\mathbf Q}$.

\subsection{QoS Maximization Under the Total Relay Power Constraint}
Now, let us consider another useful distributed beamforming
problem. Let us maximize the receiver SINR under the constraint
that the total relay transmitted power does not exceed some
maximal value $P_{\rm max}$. This problem can be written as
\begin{equation}
\max_{{\mathbf w}} \; {\rm SINR} \qquad {\rm s.t.} \quad P \leq
P_{\rm max}. \label{prob_dual_tot}
\end{equation}
Using (\ref{P_t}) and (\ref{P_s})-(\ref{P_n}), the latter problem
can be expressed as
\begin{eqnarray}
\max_{{\mathbf w}} \; \frac{{\mathbf w}^H {\mathbf Q}_s {\mathbf
w}}{{\mathbf w}^H {\mathbf Q}_i {\mathbf w} + {\mathbf w}^H
{\mathbf Q}_n {\mathbf w} +
 \sigma_{\upsilon}^2} \quad
{\rm s.t.}\quad {\mathbf w}^H {\mathbf D} {\mathbf w} \leq P_{\rm
max}. \label{prob_dual_tot_detail}
\end{eqnarray}
Introducing
\begin{equation}
\tilde {\mathbf Q}_s \triangleq {\mathbf D}^{-1/2} {\mathbf Q}_s
{\mathbf D}^{-1/2}, \quad \tilde {\mathbf Q}_{i+n} \triangleq
{\mathbf D}^{-1/2} ({\mathbf Q}_i+{\mathbf Q}_n) {\mathbf
D}^{-1/2} \nonumber
\end{equation}
we obtain that the problem (\ref{prob_dual_tot_detail}) can be
rewritten as
\begin{equation}
\max_{\tilde{\mathbf w}} \; \frac{{\tilde{\mathbf w}}^H
{\tilde{\mathbf Q}}_s {\tilde{\mathbf w}}} {{\tilde{\mathbf w}}^H
{\tilde{\mathbf Q}}_{i+n} {\tilde{\mathbf w}} +
\sigma_{\upsilon}^2} \qquad {\rm s.t.} \quad \|\tilde{\mathbf
w}\|^2 \leq P_{\rm max} \label{prob_dual_tot_tilde}
\end{equation}
where, as before, $\tilde {\mathbf w} \triangleq {\mathbf D}^{1/2}
{\mathbf w}$. It can be easily proved that the objective function
in (\ref{prob_dual_tot_tilde}) achieves its maximum when the
constraint is satisfied with equality (i.e., $\|\tilde{\mathbf
w}\|^2=P_{\rm max}$). Therefore, the problem
(\ref{prob_dual_tot_tilde}) can be rewritten as
\begin{equation}
\max_{\tilde{\mathbf w}} \; \frac{{\tilde{\mathbf w}}^H
{\tilde{\mathbf Q}}_s {\tilde{\mathbf w}}} {{\tilde{\mathbf w}}^H
({\tilde{\mathbf Q}}_{i+n} + (\sigma_{\upsilon}^2/P_{\rm max})
{\mathbf I} ) {\tilde{\mathbf w}}} \qquad {\rm s.t.} \quad
\|\tilde{\mathbf w}\|^2 = P_{\rm max}. \label{prob_dual_tot_eq}
\end{equation}

In contrast to the problem of Section~\ref{subsec1}, the problem
(\ref{prob_dual_tot_eq}) is always feasible
because for any positive $P_{\rm max}$, its
feasible set is nonempty. Using the results of \cite{sha03}
(where a mathematically similar problem has been discussed in a
different context), we conclude that the objective function in
(\ref{prob_dual_tot_eq}) is maximized when $\tilde{\mathbf w}$ is
chosen as the normalized principal eigenvector of the matrix
$({\tilde{\mathbf Q}}_{i+n} + \sigma_{\upsilon}^2/P_{\rm max}
{\mathbf I})^{-1} \tilde{\mathbf Q}_s$. Note here that any
arbitrary scaling of $\tilde{\mathbf w}$ does not change the value
of the objective function in (\ref{prob_dual_tot_eq}). However,
the so-obtained vector $\tilde{\mathbf w}$ have to be properly
scaled to satisfy the power constraint $\|\tilde{\mathbf w}\|^2 =
P_{\rm max}$. Then, the solution to (\ref{prob_dual_tot_eq}) can
be written as
\begin{equation}
\tilde{\mathbf w}_{\rm opt} = \sqrt{P_{\rm max}} {\cal P} \left\{
({\tilde{\mathbf Q}}_{i+n} + (\sigma_{\upsilon}^2/P_{\rm max})
{\mathbf I})^{-1} \tilde{\mathbf Q}_s \right\}
\end{equation}
and, therefore, the optimal beamforming weight vector and the
maximum SINR at the destination can be written as
\begin{eqnarray}
{\mathbf w}_{\rm opt} &\!\!\!=\!\!\!& \sqrt{P_{\rm max}}\,
{\mathbf D}^{-1/2} {\cal P} \left\{ ({\tilde{\mathbf Q}}_{i+n} +
(\sigma_{\upsilon}^2/P_{\rm max}) {\mathbf I})^{-1}
\tilde{\mathbf Q}_s \right\} \label{sol_w2} \\
{\rm SINR}_{\rm max} &\!\!\!=\!\!\!& {\cal L}_{\rm max}
\left\{({\tilde{\mathbf Q}}_{i+n} + (\sigma_{\upsilon}^2/P_{\rm
max}) {\mathbf I})^{-1} \tilde{\mathbf Q}_s\right\}
\label{SINR_max}
\end{eqnarray}
respectively.

\subsection{QoS Maximization Under the Individual Relay Power
Constraints}\label{QoS_ind}

Now, let us consider another relevant distributed beamforming
problem which differs from (\ref{prob_dual_tot}) is that the
individual relay power constraints are used instead of the total
relay power constraints. This problem can be written as
\begin{eqnarray}
\max_{{\mathbf w}} \; {\rm SINR} \quad {\rm s.t.} \quad p_m \leq
p_{m, {\rm max}},\quad m = 1,\cdots,R \label{prob_dual_orig}
\end{eqnarray}
where $p_{m, {\rm max}}$ denotes the maximal transmitted power of
the $m$th relay. Using (\ref{p_i_w}) and (\ref{P_s})-(\ref{P_n}),
and introducing a new auxiliary variable $\tau>0$
\cite{boyd04}, the problem (\ref{prob_dual_orig})
can be transformed to
\begin{eqnarray}
\!\!\!\!\!\!\!&\!\!\!\!\!\!\!&\max_{{\mathbf w},\tau} \quad \tau  \nonumber\\
\!\!\!\!\!\!\!&\!\!\!\!\!\!\!&{\rm s.t.} \quad \frac{{\mathbf w}^H
{\mathbf Q}_s {\mathbf w}} { {\mathbf w}^H {\mathbf Q}_i {\mathbf w} +
{\mathbf w}^H {\mathbf Q}_n {\mathbf w} + \sigma_\upsilon^2} \geq \tau^2 \label{prob_dual_tau}\\
\!\!\!\!\!\!\!&\!\!\!\!\!\!\!&\;\; \qquad {\mathbf w}^H {\mathbf D}_m {\mathbf w}
 \leq p_{m, {\rm max}},\qquad m = 1,\cdots,R. \nonumber
\end{eqnarray}
The first constraint in (\ref{prob_dual_tau}) can be rewritten as
\begin{eqnarray}
\sqrt{P_s} |{\mathbf w}^H {\mathbf h}| \geq \tau \sqrt{ {\mathbf
w}^H {\mathbf Q}_i {\mathbf w} + {\mathbf w}^H {\mathbf Q}_n
{\mathbf w} + \sigma_\upsilon^2} \label{constr_abs}
\end{eqnarray}
where ${\mathbf h} \triangleq {\mathbf A}^H {\mathbf h}_0$. We
observe that any arbitrary phase rotation of ${\mathbf w}$ will
not change the value of the objective function in
(\ref{prob_dual_tau}). Using a proper phase rotation, we have that
the constraint (\ref{constr_abs}) is equivalent to
\begin{eqnarray}
\sqrt{P_s} {\mathbf w}^H {\mathbf h} &\!\!\!\geq\!\!\!& \tau
\sqrt{ {\mathbf w}^H {\mathbf Q}_i {\mathbf w} + {\mathbf w}^H
{\mathbf Q}_n {\mathbf w} + \sigma_\upsilon^2}
\label{constr_pos}\\
{\rm Im}\{{\mathbf w}^H {\mathbf h}\} &\!\!\!\!=\!\!\!& 0
\label{constr_pos1}
\end{eqnarray}
where ${\rm Im}\{\cdot\}$ denotes the imaginary part of a complex
value. Note, however, that (\ref{constr_pos1}) can be omitted as
it is automatically taken into account in (\ref{constr_pos}) by
the fact that the right-hand side of (\ref{constr_pos}) is
non-negative. Then, the problem (\ref{prob_dual_tau}) can be
rewritten as
\begin{eqnarray}
\!\!\!\!\!\!\!&\!\!\!\!\!\!\!&\max_{{\mathbf w},\tau}
\quad \tau  \nonumber\\
\!\!\!\!\!\!\!&\!\!\!\!\!\!\!&{\rm s.t.}
\quad \sqrt{P_s} {\mathbf w}^H {\mathbf h}
\geq \tau \sqrt{ {\mathbf w}^H {\mathbf Q}_i {\mathbf w} +
{\mathbf w}^H {\mathbf Q}_n {\mathbf w} + \sigma_\upsilon^2} \label{prob_dual_sqrt}\\
\!\!\!\!\!\!\!&\!\!\!\!\!\!\!&\;\; \qquad {\mathbf w}^H {\mathbf
D}_m {\mathbf w} \leq p_{m, {\rm max}},\quad m = 1,\cdots,R.
\nonumber
\end{eqnarray}
Let
\begin{eqnarray}
{\mathbf B} &\!\!\! \triangleq \!\!\!& \left[
\begin{array}{cc} \sigma_{\upsilon}^2 & {\mathbf 0}_{RL_w \times 1}^T \\
{\mathbf 0}_{RL_w \times 1}& {\mathbf Q}_i + {\mathbf Q}_n \end{array} \right] =
{\mathbf U}^H{\mathbf U}\\
{\mathbf D}_m &\!\!\!=\!\!\!& {\mathbf V}_m^H {\mathbf V}_m,\quad m=1,\cdots,R
\end{eqnarray}
be the Cholesky factorizations of the matrices ${\mathbf B}$ and
${\mathbf D}_m$, respectively. Introducing new notations
\begin{equation}
\breve {\mathbf w} \triangleq [1,{\mathbf w}^T]^T,\quad
\breve {\mathbf V}_m \triangleq [{\mathbf 0}_{RL_w \times 1},{\mathbf V}_m], \quad
\breve {\mathbf h}  \triangleq [0,{\mathbf h}^T]^T
\end{equation}
we can further rewrite the problem (\ref{prob_dual_sqrt}) as
\begin{eqnarray}
\!\!\!\!\!\!\!&\!\!\!\!\!\!\!& \max_{\breve {\mathbf w},\tau} \quad \tau \nonumber\\
\!\!\!\!\!\!\!&\!\!\!\!\!\!\!&{\rm s.t.} \quad \sqrt{P_s} \breve{\mathbf w}^H
\breve{\mathbf h}  \geq  \tau \| {\mathbf U} \breve{\mathbf w} \|  \nonumber\\
\!\!\!\!\!\!\!&\!\!\!\!\!\!\!& \; \qquad \|{\breve{\mathbf V}}_m
{\breve{\mathbf w}}\| \leq \sqrt{p_{m, {\rm max}}},\quad m =
1,\cdots,R \label{prob_dual_socp}\\
\!\!\!\!\!\!\!&\!\!\!\!\!\!\!& \; \qquad \breve{\mathbf w}^H{\bf
e}_1=1.
 \nonumber
\end{eqnarray}

In contrast to the problem of Section~\ref{subsec1}, the problem
(\ref{prob_dual_socp}) is always feasible. This
can be directly seen from its equivalent formulation
\eqref{prob_dual_orig} whose feasible set is always nonempty.
Moreover, the problem (\ref{prob_dual_socp}) is {\it
quasi-convex} \cite{boyd04}, because for any value of $\tau$, it
reduces to the following second-order cone programming (SOCP)
feasibility problem:
\begin{eqnarray}
\!\!\!\!\!\!\!&\!\!\!\!\!\!\!& {\rm find} \quad \breve{\mathbf w} \nonumber\\
\!\!\!\!\!\!\!&\!\!\!\!\!\!\!&{\rm s.t.} \quad \sqrt{P_s} \breve{\mathbf w}^H \tilde{\mathbf h}
\geq  \tau \|{\mathbf U} \breve{\mathbf w} \| \nonumber \\
\!\!\!\!\!\!\!&\!\!\!\!\!\!\!& \; \qquad \|{\breve{\mathbf V}}_m
{\breve{\mathbf w}}\| \leq \sqrt{p_{m,{\rm max}}},\quad m =
1,\cdots,R \label{prob_dual_socp_feas}\\
\!\!\!\!\!\!\!&\!\!\!\!\!\!\!& \; \qquad \breve{\mathbf w}^H {\bf
e}_1=1.\nonumber
\end{eqnarray}

Let $\tau_{*}$ be the optimal value of $\tau$ in
(\ref{prob_dual_socp}). Then, for any $\tau>\tau_{*}$, the problem
(\ref{prob_dual_socp_feas}) is infeasible. On the contrary, if
(\ref{prob_dual_socp_feas}) is feasible, then we conclude that
$\tau \leq \tau_{*}$. Hence, the optimum $\tau_{*}$ and the
optimal weight vector $\breve{\mathbf w}_{*}$ can be found using
the bisection search technique discussed in \cite{havary08}.
Assuming that $\tau_{*}$ lies in the interval $[\tau_l,\tau_u]$,
the bisection search procedure to solve (\ref{prob_dual_socp}) can
be summarized as the following sequence of steps:
\begin{enumerate}
\item $\tau := (\tau_l+\tau_u)/2$. \label{tau=middle}
\item Solve the convex feasibility problem (\ref{prob_dual_socp_feas}).
If (\ref{prob_dual_socp_feas}) is feasible, then $\tau_l:=\tau$,
otherwise $\tau_u:=\tau$.
\item If $(\tau_u-\tau_l)< \varepsilon$ then stop. Otherwise, go to
Step \ref{tau=middle}. \label{stop}
\end{enumerate}
Here, $\varepsilon$ is the error tolerance value in $\tau$.

Note that the feasibility problem (\ref{prob_dual_socp_feas}) is a
standard SOCP problem, which can be efficiently solved using
interior point methods \cite{sturm99} with the worst-case
complexity of ${\cal O}((RL_w)^{3.5})$. The initial interval for
the bisection search can be selected as
$[\tau_l,\tau_u]=[0,\sqrt{{\rm SINR}_{\rm max}(P_{\rm max})}]$,
where ${\rm SINR}_{\rm max}(P_{\rm max})$ can be obtained from
(\ref{SINR_max}) by choosing $P_{\rm max}=\sum_{m=1}^R p_{m, {\rm
max}}$. This particular choice is motivated by the fact that the
optimal SINR of (\ref{prob_dual_tot}) always upper bounds the
optimal SINR of (\ref{prob_dual_orig}).

{\it Remark:} It is worth noting that the total
power constraint can be easily added to
\eqref{prob_dual_socp_feas} just as one more second-order cone
constraint
$$
\|{\bf V}{\bf w}\|\le \sqrt{P_{\rm max}}
$$
where ${\bf V}^H{\bf V}$ is the Cholesky factorization of ${\bf
D}$. This allows us to directly extend the approach of
\eqref{prob_dual_socp_feas} to a practically important case when
both the individual and the total power constraints have to be
taken into account \cite{zheng}.

\subsection{Relationships Between the Proposed Methods and Earlier Techniques in
the Flat Fading Case}

Let us now explore the relationship between the proposed three
methods and the techniques of \cite{jing07}, \cite{zheng} and
\cite{havary08} in the specific case when all the channels are
frequency flat and each relay filter is just a complex coefficient
($L_f=L_g=L_w=1$). In the latter case, the transmitted power of
the $m$th relay in \eqref{p_i_w} can be simplified to
\begin{eqnarray}
 p_m &\!\!\!\!=\!\!\!\!& P_s {\mathbf w}_0^H {\mathbf E}_m {\mathbf f}_0
 {\mathbf f}_0^H {\mathbf E}_m^H
 {\mathbf w}_0 + \sigma_{\eta}^2 {\mathbf w}_0^H {\mathbf E}_m
 {\mathbf E}_m^H {\mathbf w}_0\nonumber \\
&\!\!\!\!=\!\!\!\!& \sigma_{f,m}^2 |w_{0,m}|^2 + \sigma_{\eta}^2\, |w_{0,m}|^2 \label{p_m_flat}
\end{eqnarray}
where $\sigma_{f,m}^2 \triangleq P_s |f_{0,m}|^2$. Then, the total
relay transmitted power can be expressed as
\begin{eqnarray}
 P &\!\!\!\!=\!\!\!\!& \sum_{m=1}^{R} p_m = {\mathbf w}_0^H {\mathbf D}_0 {\mathbf w}_0 \label{Pt_flat}
\end{eqnarray}
where $${\mathbf D}_0 \triangleq {\rm diag}
\{\sigma_{f,1}^2,\cdots,\sigma_{f,R}^2 \} + \sigma_{\eta}^2 \,
{\mathbf I}_R.$$

The received signal at the destination (\ref{dest_sig_temp2}) can
be simplified to
\begin{eqnarray}
y(n) &\!\!\!\!=\!\!\!\!& \underbrace{ {\mathbf w}_0^H ({\mathbf
f}_0 \odot {\mathbf g}_0) s(n) }_{ \rm signal} + \underbrace{
{\mathbf w}_0^H ({\mathbf g}_0 \odot {\boldsymbol \eta}(n)) +
\upsilon (n) }_{\rm noise}.\ \ \ \quad \label{dest_sig_flat}
\end{eqnarray}
As the channel is frequency flat, there is no ISI term in
(\ref{dest_sig_flat}). Hence, the SINR reduces to SNR, and it can
be written as
\begin{eqnarray}
 {\rm SNR} = \frac{{\mathbf w}_0^H {\mathbf Q}_{s0} {\mathbf
w}_0}{{\mathbf w}_0^H {\mathbf Q}_{n0} {\mathbf w}_0 +
\sigma_{\upsilon}^2} \label{SNR_flat}
\end{eqnarray}
where
\begin{eqnarray*}
 {\mathbf Q}_{s0} &\!\!\!\! \triangleq \!\!\!\!& P_s ({\mathbf f}_0 \odot {\mathbf g}_0) ({\mathbf f}_0 \odot {\mathbf g}_0)^H = P_s {\mathbf h}_0 {\mathbf h}_0^H\\
 {\mathbf Q}_{n0} &\!\!\!\! \triangleq \!\!\!\!& \sigma_{\eta}^2 {\rm diag} \{|g_{0,1}|^2,\cdots,
 |g_{0,R}|^2 \}.
\end{eqnarray*}

Introducing the variables $0\leq \alpha_m\leq 1$ ($m=1,\cdots,R$)
and using them to express the relay powers as
\begin{equation}
p_m=\alpha_m^2 p_{m,{\rm max}} \label{p_alpha}
\end{equation}
we obtain from \eqref{p_m_flat} and \eqref{p_alpha} that
\begin{equation}
 |w_{0,m}|=\alpha_m\sqrt{\frac{p_{m,{\rm max}}}{\sigma_{\eta}^2+P_s|f_{0,m}|^2}}. \nonumber
\end{equation}
In \cite{jing07}, it is proposed to compensate the phases caused
by the transmitter-to-relay and relay-to-destination channels by a
proper choice of the phase of each $w_{0,m}$. This gives
\begin{equation}
 w_{0,m}=\alpha_m\sqrt{\frac{p_{m,{\rm max}}}{\sigma_{\eta}^2+P_s|f_{0,m}|^2}}
 \, e^{j\theta_m} \label{w_m}
\end{equation}
where $\theta_m={\rm arg}\, f_{0,m}+{\rm arg}\, g_{0,m}$.
Inserting \eqref{w_m} into \eqref{SNR_flat}, we have
\begin{equation}
 {\rm SNR} = \frac{P_s \left(\sum_{m=1}^R \alpha_m |f_{0,m}g_{0,m}| \sqrt{\frac{p_{m,{\rm max}}} {\sigma_{\eta}^2+P_s |f_{0,m}|^2} }\right)^2}
{\sigma_{\upsilon}^2+\sum_{m=1}^R \frac{\alpha_m^2 p_{m,{\rm max}}
|g_{0,m}|^2 \sigma_{\eta}^2}{\sigma_{\eta}^2 + P_s |f_{0,m}|^2}}.
\label{SNR_alpha}
\end{equation}
{From} \eqref{SNR_alpha}, it can be seen that our problem
\eqref{prob_dual_orig} in the considered particular case is
equivalent to
\begin{eqnarray}
\!\!\!\!\!\!\!&\!\!\!\!\!\!\!& \max_{\alpha_1,\cdots,\alpha_R}
\quad \frac{P_s \left(\sum_{m=1}^R \alpha_m |f_{0,m}g_{0,m}|
\sqrt{\frac{p_{m,{\rm max}}} {\sigma_{\eta}^2+P_s |f_{0,m}|^2}
}\right)^2}
{\sigma_{\upsilon}^2+\sum_{m=1}^R \frac{\alpha_m^2 p_{m,{\rm max}}
|g_{0,m}|^2 \sigma_{\eta}^2}{\sigma_{\eta}^2 + P_s |f_{0,m}|^2}} \nonumber\\
\!\!\!\!\!\!\!&\!\!\!\!\!\!\!&\quad{\rm s.t.} \qquad 0\leq
\alpha_1,\cdots,\alpha_R \leq 1 \,. \label{prob_alpha}
\end{eqnarray}
It can be readily verified that \eqref{prob_alpha} and the problem
in \cite{jing07} are identical. Moreover, the problem of
\cite{zheng} extends that of \cite{jing07} to the case when both
the individual and the total power constraints are used.
Therefore, in the flat fading case where the AF strategy is used
instead of the FF one, our approach of Section \ref{QoS_ind}
reduces to that of \cite{jing07} and, with the additional total
power constraint added to \eqref{prob_dual_socp_feas}, it reduces
to that of \cite{zheng}.

To understand the relationship of our three approaches of Section
\ref{formulation} and the techniques of \cite{havary08}, we note
that in the flat fading AF case the only difference between the
problem formulations in \cite{havary08} and our problem
formulations is that an extra statistical expectation over all the
random transmitter-to-relay and relay-to-destination channels has
been used in \cite{havary08}. Hence, in the flat fading case, the
problem formulations of \cite{havary08} transfer to our problems
\eqref{prob_orig}, \eqref{prob_dual_tot} and
\eqref{prob_dual_orig} when the instantaneous instead of the
second-order CSI is used in the methods of \cite{havary08} and
when the FF strategy is replaced by the AF one in our techniques.

\section{Simulation results}
In our simulations, we consider a relay network with $R=10$ relays
and quasi-static frequency selective transmitter-to-relay and
relay-to-destination channels with the lengths $L_f=L_g=5$.
The transmitter uses the binary phase shift keying
(BPSK) modulation. The channel impulse response coefficients are
modeled as zero-mean complex Gaussian random variables with an
exponential power delay profile \cite{Rappaport96}
\begin{eqnarray}
p(t) &=& \frac{P_R}{\sigma_t}\sum_{l=0}^{L_x} e^{-{t}/{\sigma_t}}
\delta(t-lT_s)
\end{eqnarray}
where $L_x \in \{L_f,L_g\}$, $T_s$ is the symbol duration,
$\delta(\cdot)$ is the Dirac delta function, $P_R$ is the average
power of the multipath components, and $\sigma_t$ characterizes
the delay spread. In our simulations, $P_R=1$ and $\sigma_t=2T_s$
are used. The relay and destination noises are assumed to have the
same powers, and the source transmitted power is $10$ dB higher
than the noise power. To obtain the bit error rate
(BER) curves, it has been assumed that the symbol-by-symbol
maximum likelihood (ML) decoder is used at the receiver.

In the first example, we test the approach of (\ref{sol_w}) which
is based on minimizing the total transmit power subject to the QoS
constraint. Fig.~\ref{PvsSINR} displays the total relay
transmitted power versus the minimum required SINR at the
destination for different lengths of the relay filters.
As for randomly generated signal, noise and
channel values the feasibility of (\ref{prob_ineq}) is a random
event, this problem can be infeasible for some percentage of
simulation runs. To deal with this fact in our first example, we
call the problem {\it ergodically infeasible} when the number of
simulation runs leading to infeasibility of (\ref{prob_ineq}) is
larger than the half of the total number of simulation runs;
otherwise, this problem is classified as {\it ergodically
feasible}. If the problem is ergodically infeasible, the
corresponding points are dropped from the figures displaying the
behaviour of the total transmitted power. In the case of ergodic
feasibility, the corresponding points are computed by averaging
over the ``feasible'' runs and displayed in these figures. For
example, there are several dropped points in Fig.~\ref{PvsSINR} at
high SINR values that correspond to the case of ergodic
infeasibility of (\ref{prob_ineq}). To further illustrate the
effects of the required SINR and $L_w$ on the feasibility of the
problem (\ref{prob_ineq}), the probability that this problem is
feasible is displayed in Fig.~\ref{INFvsSINR}. The latter
probability is referred to as {\it feasibility probability}.

It can be seen from Fig.~\ref{PvsSINR} that using the FF strategy
at the relays, one can substantially reduce the total relay
transmitted power as compared to the AF approach.
Also, from Figs.~\ref{PvsSINR} and \ref{INFvsSINR}
is is clear that the FF strategy significantly improves the SINR
feasibility range of the considered distributed beamforming
problem. These improvements are monotonic in $L_w$:
for example, according to Fig.~\ref{PvsSINR}, for
${\rm SINR}=12$~dB this problem is ergodically infeasible for the
relay filter lengths $L_w=1,2$, but it becomes ergodically
feasible for any $L_w\geq 3$. These observations are further
supported by Fig.~\ref{INFvsSINR} that demonstrates that the
feasibility probability can be substantially improved by
increasing the relay filter length $L_w$.

Figs.~\ref{PvsL} and \ref{INFvsL} depict the total
relay transmitted power and feasibility probability versus the
relay filter length $L_w$ for different values of the required
SINR at the destination. Similarly to the previous
two figures, Figs.~\ref{PvsL} and \ref{INFvsL} clearly
demonstrate that the performance (in terms of the relay
transmitted power) and feasibility of the QoS constraint can be
substantially improved by using the FF approach in lieu of the AF
strategy, and these improvements become more pronounced when
increasing the relay filter length. Note that
theoretically, to fully compensate the effect of frequency
selective source-to-relay and relay-to-destination channels, it is
required that $L_w\ge L_f+L_g-1$. However, in the exponential
power delay profile case (where these channels are mainly
determined by several first taps), they are well compensated even
with $L_w<L_f+L_g-1$. As follows from Fig.~\ref{PvsL}, depending
on the SINR value, the filter lengths $L_w=2$ to $L_w=5$ appear to
be sufficient.

In the second example, we test the approach of (\ref{sol_w2})
which maximizes the QoS subject to the total power constraint.
Note that the problem (\ref{prob_dual_tot_eq}) is
always feasible and, therefore, there is no infeasibility issue in
this example. Fig.~\ref{SINRvsP_tot} shows the achieved SINR
versus the total relay transmitted power $P_{\rm max}$ for
different lengths of the relay filters. Fig.~\ref{SINRvsL_tot}
depicts the SINR versus the relay filter length $L_w$ for
different values of the total relay transmitted power $P_{\rm
max}$. It can be seen from these figures that the QoS can be
substantially improved by increasing the filter length $L_w$.

To illustrate the receiver error probability
performance of the FF relaying approach based on the particular
example of the distributed beamformer (\ref{sol_w2}), in
Figs.~\ref{SERvsP_tot} and \ref{SERvsL_tot} we display the
receiver BERs versus the total transmitted power $P_{\rm max}$ and
the relay filter length $L_w$, respectively. It can be observed
from these two figures that increasing the filter length, we can
substantially decrease the receiver error probability.

In our last example, the approach of Section \ref{QoS_ind} is
tested, which maximizes the QoS under the individual relay power
constraints using a combination of (\ref{prob_dual_socp_feas}) and
bisection search. As in the previous example, the
underlying problem (\ref{prob_dual_socp}) is always feasible and,
therefore, there is no infeasibility issue here. It is assumed
that all the relays have the same maximal allowed transmitted
power $p_{\rm max}$. Fig.~\ref{SINRvsP} displays the SINR versus
$p_{\max}$ for different lengths of the relay filters.
Fig.~\ref{SINRvsL} shows the SINR versus the relay filter length
$L_w$ for different values of $p_{\rm max}$.

The conclusions following from Figs.~\ref{SINRvsP} and
\ref{SINRvsL} are quite similar to that following from
Figs.~\ref{SINRvsP_tot} and \ref{SINRvsL_tot}. As the individual
power constraints are tighter than the total one, the SINRs
achieved for any value of the total transmitted power in
Figs.~\ref{SINRvsP} and \ref{SINRvsL} are a few dB's lower than
that achieved in Figs.~\ref{SINRvsP_tot} and \ref{SINRvsL_tot}
for the same value of $Rp_{\rm max}$.

Summarizing, all our examples clearly verify that the proposed FF
strategy substantially outperforms the AF approach in the
frequency selective fading case.

\section{Conclusion}
The problem of distributed network beamforming has been addressed
in the case when the transmitter-to-relay and relay-to-destination
channels are frequency selective. To compensate for the effects of
these channels, a novel filter-and-forward relay beamforming
strategy has been proposed as an extension of the traditional
amplify-and-forward protocol. According to the former strategy,
FIR filters have to be used at the relay nodes to remove ISI.

Three relevant half-duplex filter-and-forward beamforming problems
have been formulated and solved. Our first technique minimizes the
total relay transmitted power subject to the destination QoS
constraint, whereas the second and third methods maximize the
destination QoS subject to the total and individual relay
transmitted power constraints, respectively. For the first and
second approaches, closed-form beamformers have been obtained,
whereas the third beamformer is computed using convex
optimization, via a combination of bisection search and
second-order cone programming. It has been also shown that the
latter convex optimization-based relay beamformer can be easily
extended to the practically important case when the individual and
total power constraints should be jointly taken into account.

Our simulation results demonstrate that in the frequency selective
fading case, the proposed filter-and-forward beamforming strategy
provides substantial performance improvements as compared to the
commonly used amplify-and-forward relay beamforming approach.

\newpage

\begin{figure}
\centering
\resizebox{12.0cm}{!}{\includegraphics{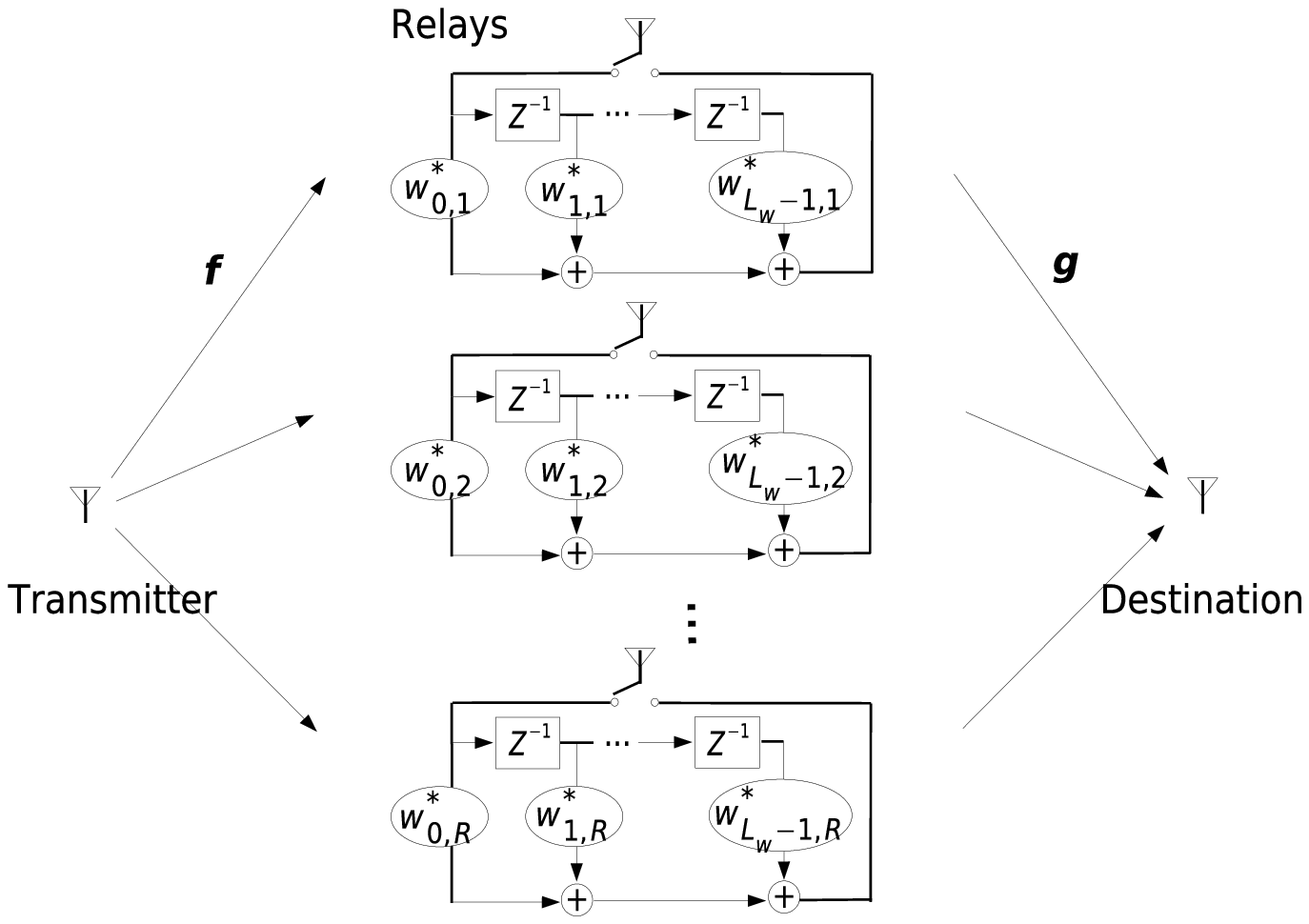}}
\caption{Filter-and-forward relay network.} \label{sys_model}
\end{figure}

\begin{figure}
\centering \resizebox{8.5cm}{!}{\includegraphics{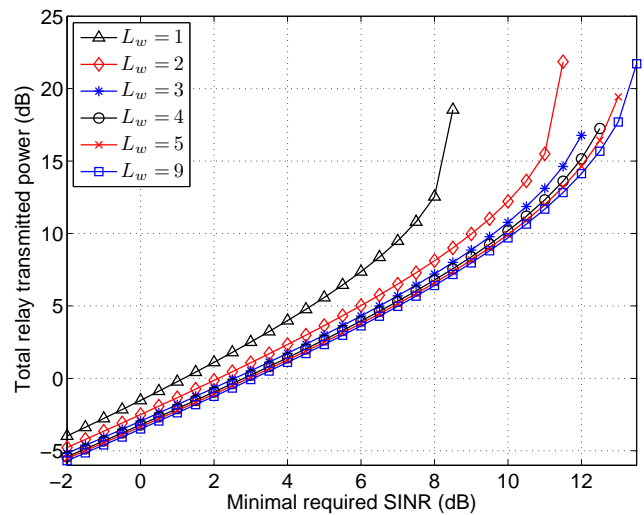}}
\vspace{-0.3cm} \caption{Total relay transmitted power versus
required SINR; first example.} \label{PvsSINR}
\end{figure}

\begin{figure}
\centering \resizebox{8.5cm}{!}{\includegraphics{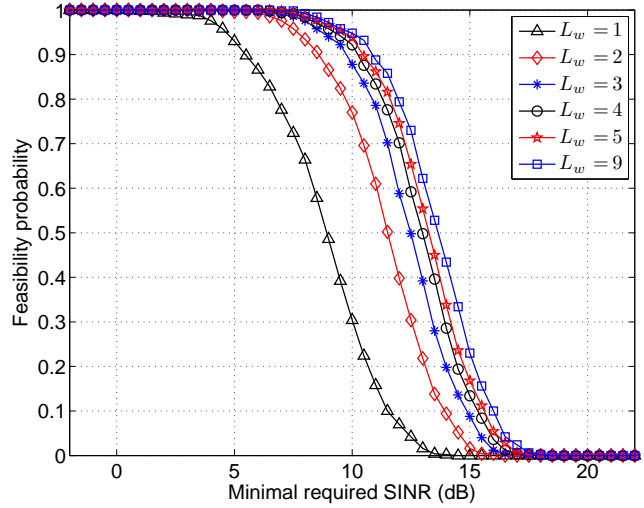}}
\vspace{-0.3cm} \caption{Feasibility probability of the problem
(\ref{prob_ineq}) versus required SINR; first example.}
\label{INFvsSINR}
\end{figure}

\begin{figure}
\centering
\resizebox{8.5cm}{!}{\includegraphics{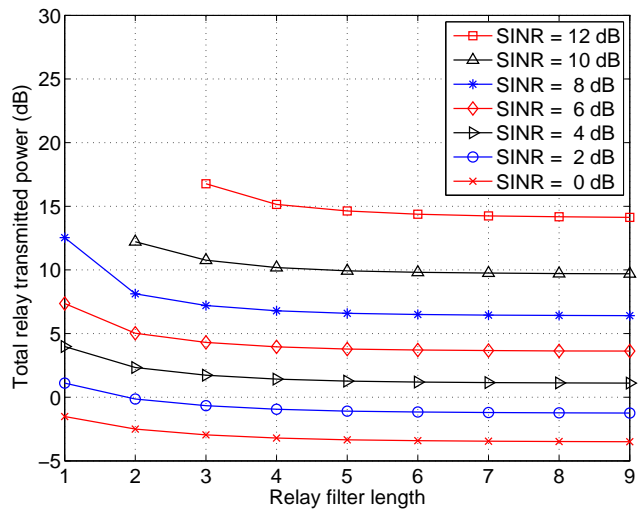}}\vspace{-0.3cm}
\caption{Total relay transmitted power versus relay filter length
$L_w$; first example.} \label{PvsL}
\end{figure}

\begin{figure}
\centering
\resizebox{8.5cm}{!}{\includegraphics{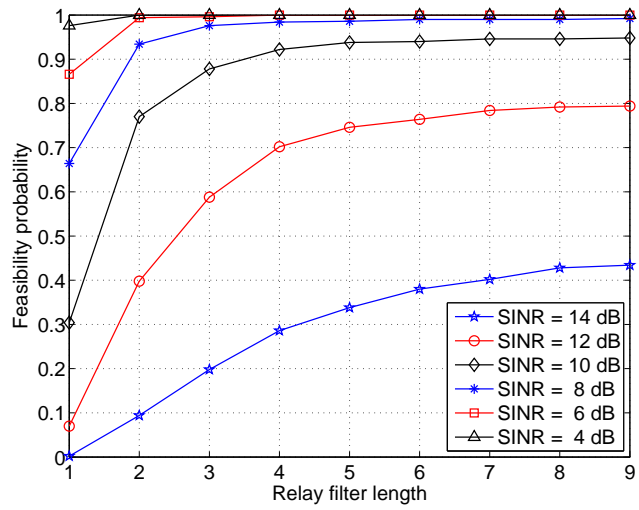}}\vspace{-0.3cm}
\caption{Feasibility probability of the problem (\ref{prob_ineq})
versus relay filter length $L_w$; first example.} \label{INFvsL}
\end{figure}

\begin{figure}
\centering \resizebox{8.5cm}{!}{\includegraphics{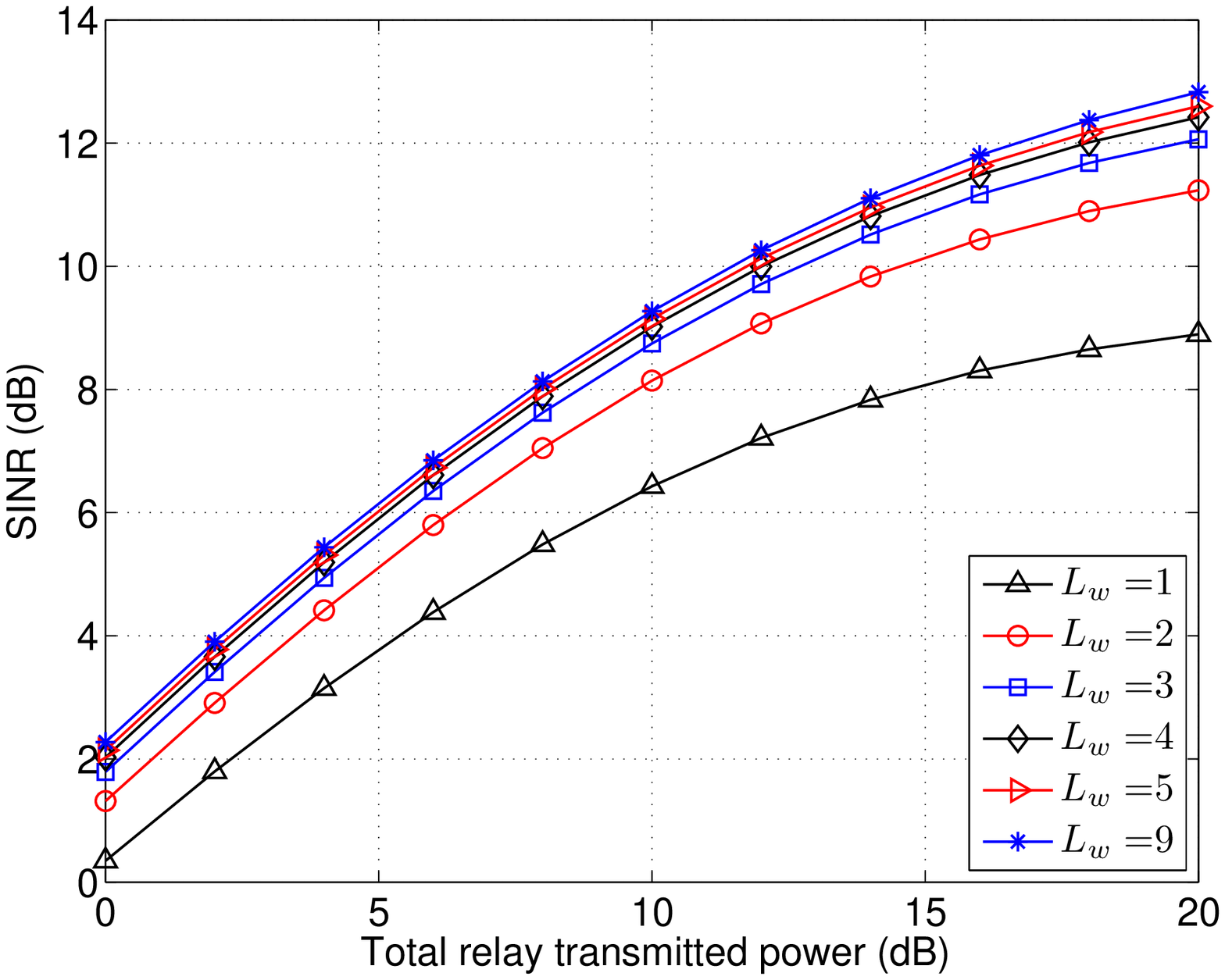}}
\caption{SINR versus the maximal total relay transmitted power
$P_{\rm max}$; second example.} \label{SINRvsP_tot}
\end{figure}

\begin{figure}
\centering \resizebox{8.5cm}{!}{\includegraphics{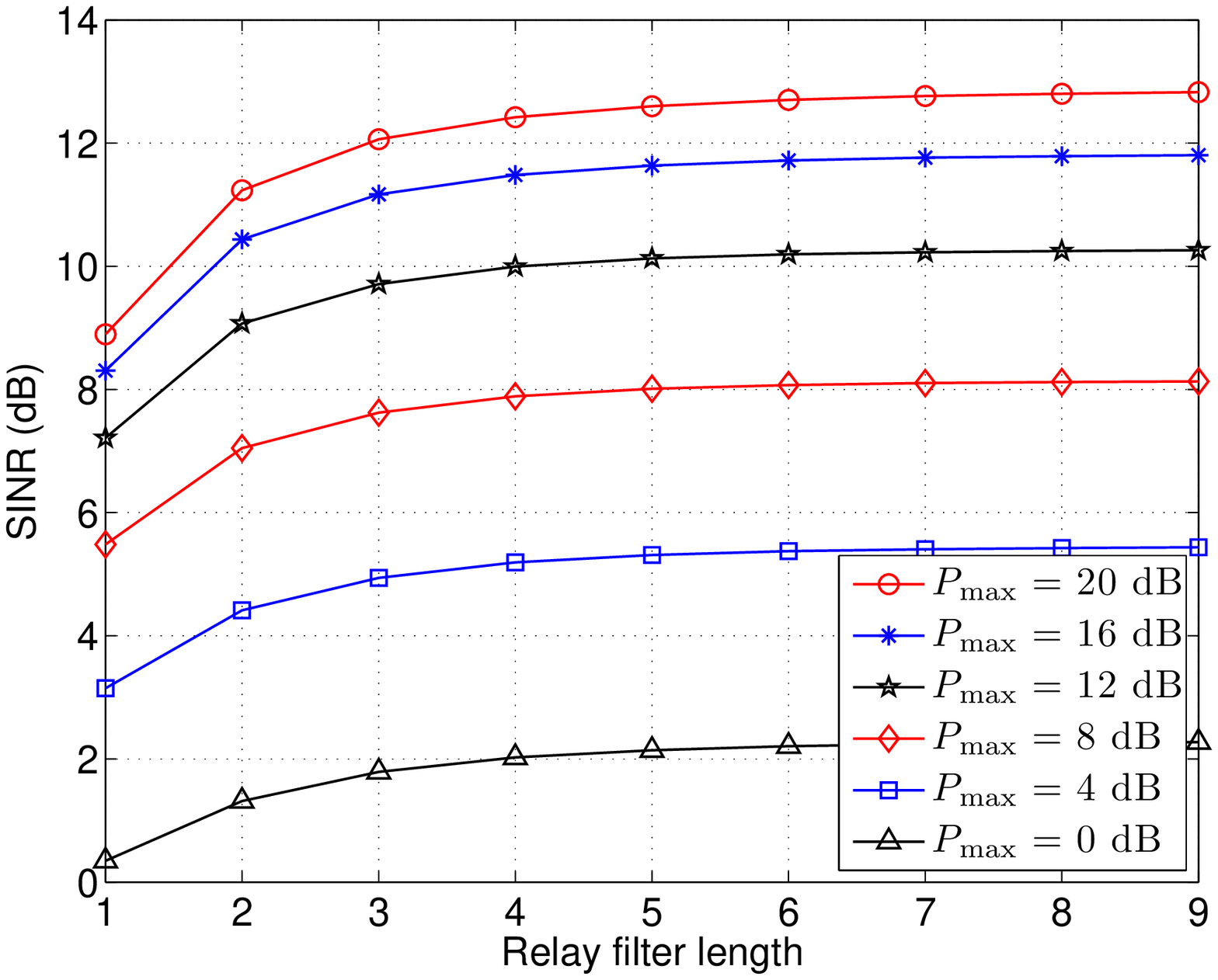}}
\caption{SINR versus relay filter length $L_w$; second example.}
\label{SINRvsL_tot}
\end{figure}

\begin{figure}
\centering \resizebox{8.5cm}{!}{\includegraphics{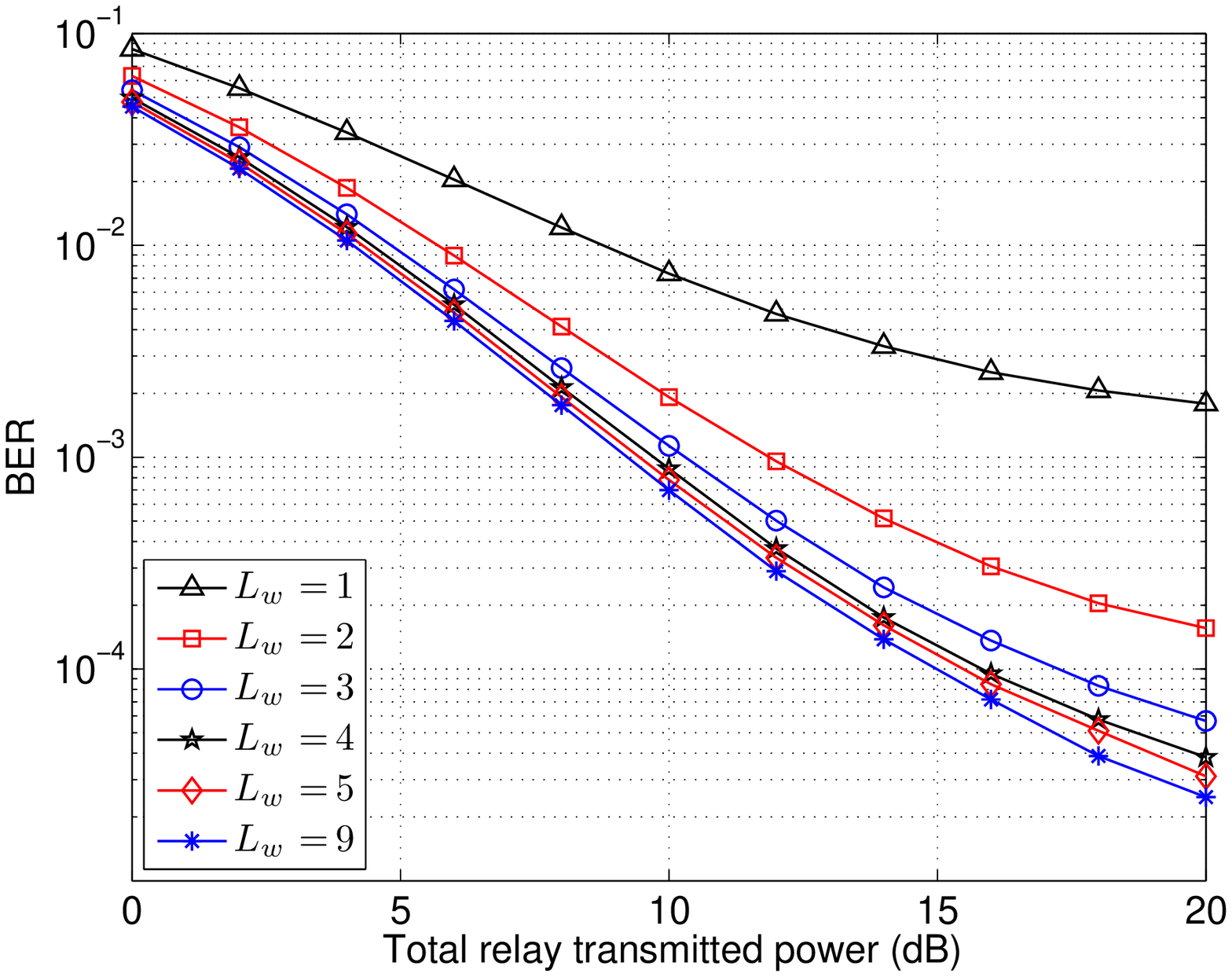}}
\caption{SER versus the maximal total relay transmitted power
$P_{\rm max}$; second example.} \label{SERvsP_tot}
\end{figure}

\begin{figure}
\centering \resizebox{8.5cm}{!}{\includegraphics{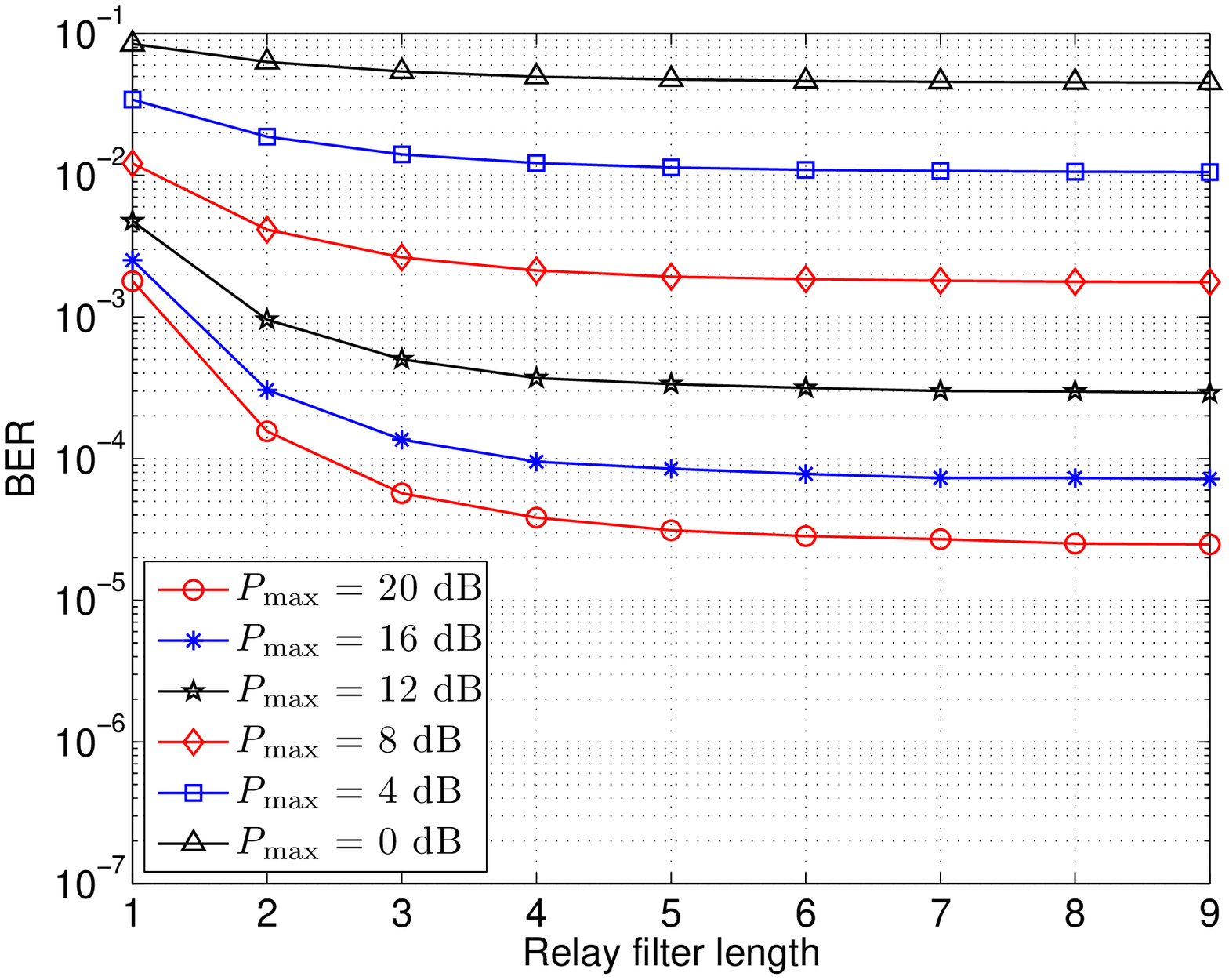}}
\caption{SER versus relay filter length $L_w$; second example.}
\label{SERvsL_tot}
\end{figure}

\begin{figure}
\centering \resizebox{8.5cm}{!}{\includegraphics{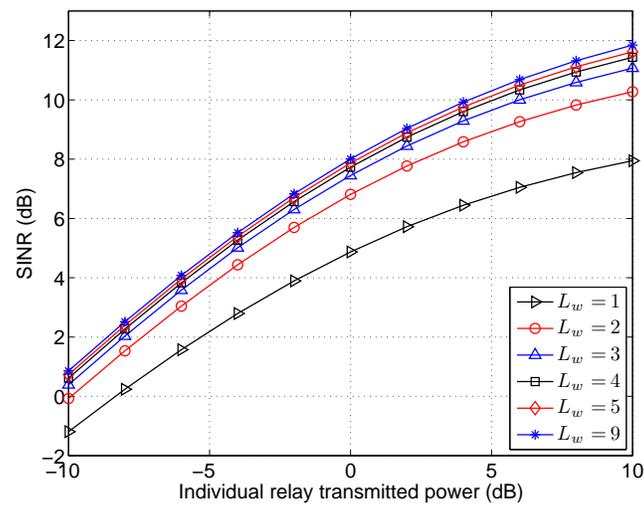}}
\caption{SINR versus the maximal individual relay transmitted
power $p_{\rm max}$; third example.} \label{SINRvsP}
\end{figure}

\begin{figure}
\centering \resizebox{8.5cm}{!}{\includegraphics{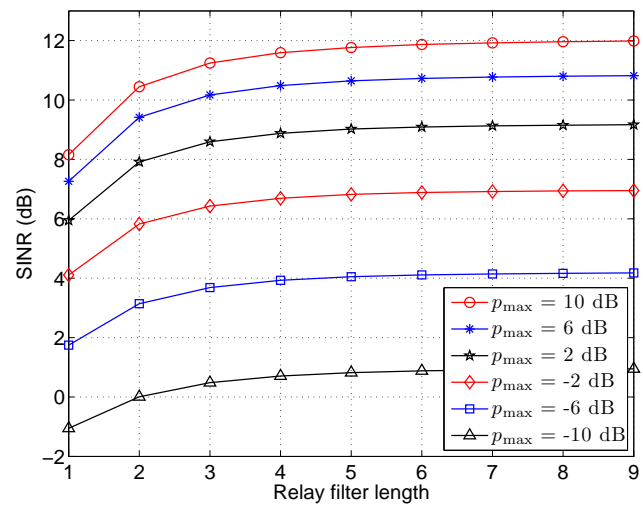}}
\caption{SINR versus relay filter length $L_w$; third example.}
\label{SINRvsL}
\end{figure}

\end{document}